\documentclass[a4paper,onecolumn,superscriptaddress,11pt,accepted=2021-02-18]{quantumarticle}
\pdfoutput=1
\usepackage[utf8]{inputenc}
\usepackage[english]{babel}
\usepackage[T1]{fontenc}
\usepackage{amsmath}
\usepackage{amssymb}
\usepackage{hyperref}
\usepackage{qcircuit}
\usepackage{xcolor}
\usepackage{graphicx}
\usepackage[numbers,sort&compress]{natbib}
\usepackage{tikz}

\newcommand{\sinc}{\mathrm{sinc}}

\begin{document}

\title{A quantum algorithm for the direct estimation of the steady state of open quantum systems}
\date{\today}
\author{Nathan Ramusat}
\affiliation{Institute of Physics, Ecole Polytechnique Fédérale de Lausanne (EPFL), CH-1015 Lausanne, Switzerland}
\author{Vincenzo Savona}
\affiliation{Institute of Physics, Ecole Polytechnique Fédérale de Lausanne (EPFL), CH-1015 Lausanne, Switzerland}
\email{vincenzo.savona@epfl.ch}
%\homepage{http://quantum-journal.org}
%\orcid{0000-0003-0290-4698}
%\thanks{You can use the \texttt{\textbackslash{}email}, \texttt{\textbackslash{}homepage}, and \texttt{\textbackslash{}thanks} commands to add additional information for the preceding \texttt{\textbackslash{}author}. If applicable, this can also be used to indicate that a work has previously been published in conference proceedings.}

\maketitle

\begin{abstract}
Simulating the dynamics and the non-equilibrium steady state of an open quantum system are hard computational tasks on conventional computers. For the simulation of the time evolution, several efficient quantum algorithms have recently been developed. However, computing the non-equilibrium steady state as the long-time limit of the system dynamics is often not a viable solution, because of exceedingly long transient features or strong quantum correlations in the dynamics. Here, we develop an efficient quantum algorithm for the direct {\color{black} estimation of averaged expectation values of observables on} the non-equilibrium steady state, {\color{black} thus bypassing the time integration of the master equation}. The algorithm {\color{black} encodes the vectorized representation of the density matrix on a quantum register,} and makes use of quantum phase estimation to approximate the eigenvector associated to the zero eigenvalue of the generator of the system dynamics. We show that the output state of the algorithm allows to estimate expectation values of observables on the steady state. Away from critical points, where the Liouvillian gap scales as a power law of the system size, the quantum algorithm performs with exponential advantage compared to exact diagonalization.
\end{abstract}

\section{Introduction}

Open quantum systems are rapidly emerging as a major research field \cite{Carusotto2013,Noh2017a,Hartmann2016}. On the fundamental side, they can display new classes of universal physical properties such as dissipative phase transitions and topological phases, while applications can span from novel paradigms of quantum simulators \cite{Rota2019,Rota2019a} to the accurate modeling of noise in modern quantum computing platforms \cite{Michael2016}. The numerical simulation of the dynamics of a many-body open quantum system is generally a hard computational task, as it combines the unitary evolution generated by the many-body Hamiltonian with the non-unitary evolution induced by the interaction with the environment. Similarly, computing the non-equilibrium steady state (NESS) -- i.e. the state reached asymptotically in the long-time limit -- is the computational analog of simulating the ground state of a many-body closed quantum system, and generally embodies an analogous computational challenge.

Current approaches \cite{Weimer2019} to the solution of the Lindblad-Von-Neumann master equation for the density matrix \cite{Breuer2002} -- describing the case of a Markovian environment -- include in addition to the exact numerical solution of the resulting differential equations, tensor-network methods \cite{Werner2016,Verstraete2004,Zwolak2004}, projector \cite{Nagy2018} and time-dependent variational \cite{Nagy2019,Hartmann2019,Vicentini2019,Yoshioka2019} quantum Monte Carlo, various quantum trajectory approaches, as well as methods introducing various levels of approximation \cite{Biella2018,Casteels2018,Verstraelen2020,Verstraelen2018,Lange2018}. When the goal of the simulation is to know the NESS, integrating the system dynamics over long times is generally not an optimal strategy. In presence of critical slowing down, the time required to actually reach the NESS may be prohibitively large, and the transient dynamics may explore highly quantum correlated states -- generally more difficult to simulate -- even when the quantum correlations in the NESS are moderate. More specific approaches aiming directly at the simulation of the NESS without simulating the underlying dynamics have been developed, either in terms of variational tensor network \cite{Kshetrimayum2017,Mascarenhas2015,Cui2015}, and specific real-space decimation schemes \cite{Finazzi2015}. The runtime of most of these approaches scales exponentially with the number of degrees of freedom, while approximate methods display power-law scaling but at the cost of a limited predictive power when in presence of significant quantum correlations in the system.

Quantum algorithms for the simulation of open quantum systems by computing the system dynamics have recently been proposed \cite{Chenu2017,Wang2011,Barreiro2011,Kliesch2011,GarciaPerez2020,Su2020,Hu2020,Sweke2014,Sweke2015,DiCandia2015,Sweke2016,Childs2017,Cleve2017,Liu2020}. The time-evolution of the Lindblad-Von-Neumann master equation is non-unitary, hence not suited for a direct implementation as a digital quantum simulation. While some special cases of system-environment interaction have been shown to translate into a unitary stochastic quantum evolution \cite{Chenu2017}, more generally the non-unitary dynamics is mapped onto a unitary dynamics on an appropriately dilated Hilbert space, which models an effective environment. Similarly, a time evolution quantum algorithm was proposed to compute the thermal equilibrium state of a system in interaction with a thermal bath \cite{Terhal2000}. Recently, a hybrid quantum algorithm has been proposed for the estimation of the NESS, based on a variational ansatz for the density matrix \cite{Yoshioka2019a}. 

{\color{black}Here, we propose a different approach consisting in the direct estimation of the averaged expectation values of observables on the NESS, without requiring the integration of the system dynamics. This is achieved by representing the density matrix in vectorized form and mapping the steady-state condition onto a linear system of equations. The quantum algorithm then uses quantum phase estimation (QPE) and the HHL algorithm \cite{Harrow2009} to solve the linear system. The output state is an estimate of the elements of the density matrix of the NESS, encoded in a vectorized representation onto the output state of a quantum register. The initial state is prepared leveraging the known spectral properties of the generator of the open-system dynamics, so to obtain a large overlap with the target output state. In this way, the success probability of a single QPE run is $O(1)$. It is shown that in cases in which the Liouvillian gap -- i.e. the asymptotic relaxation rate towards the NESS -- decreases as a power law of the system size, the algorithm runs with exponential speedup compared to exact diagonalization of the master equation on classical hardware. By using established methods for the quantum measurement of expectation values \cite{Knill2007}, is it finally shown how to directly estimate averages of the expectation value of physical observables on the NESS statistical ensemble.}

The article is organized as follows. In Section 2 the master equation formalism is introduced and cast in a form suitable to be encoded on a quantum computer. In Section 3 the quantum algorithm is presented and an analysis of the success probability, gate cost and errors is performed. In Section 4 the formalism to estimate expectation values on the NESS is derived, and its accuracy characterized. Section 5 contains a numerical analysis of the algorithm applied to a simple driven-dissipative problem. The conclusions are presented in Section 6. Appendix A develops the oracular part of the algorithm for the specific case of a dissipative transverse Ising spin model.

\section{Formalism}

The dynamics of an open quantum system interacting with a Markovian environment is described by the Lindblad-Von Neumann master equation $\dot{\hat\rho}={\cal L}(\hat\rho)$ \cite{Breuer2002}, where $\hat\rho$ is the density operator, and the Liouvillian super-operator is expressed as (setting $\hbar=1$)
\begin{equation}
{\cal L}(\hat\rho)=-i\left[\hat H,\hat\rho\right]-\frac{1}{2}\sum_j\left(\left\{\hat A_j^\dagger\hat A_j,\hat\rho\right\}-2\hat A_j\hat\rho\hat A_j^\dagger\right)\,.
\label{Lindblad}
\end{equation}
Here, $\hat H$ is the system Hamiltonian and $\hat A_j$ are jump operators describing the transitions induced on the system by the environment. For the scope of this work, we will assume that both $\hat H$ and the $\hat A_j$'s are quasi-local operators, i.e. they are expressed as sums of tensor products of at most few local degrees of freedom of the system, so that they can be efficiently implemented in a quantum circuit \cite{Lloyd1996}. The case of global jump operators, describing transitions between eigenstates of $\hat H$, can also be addressed, provided an efficient way of approximating these operators by a sequence of quantum gates exists. We assume that a unique non-equilibrium steady state (NESS) exists and is therefore characterized by the condition ${\cal L}(\hat\rho_{ss})=0$. The existence and uniqueness of the NESS can be established within rather general assumptions, and is in particular verified for problems defined in a finite-dimensional Hilbert space \cite{Nigro2019}, as it will always be the case on a quantum circuit implementation of the problem.

{\color{black} In order to encode the density matrix onto the state of a quantum register, we adopt the vectorized representation, whereby operators on the Hilbert space ${\cal H}$ are mapped onto vectors in the tensor product space ${\cal H}\otimes{\cal H}$.} Given an orthonormal basis, a density operator $\hat\rho=\sum_{jk}\rho_{jk}|j\rangle\langle k|$ maps onto the vector $|\rho\rangle=\sum_{jk}\rho_{jk}|j\rangle\otimes|k\rangle$ (appropriately normalized). This essentially corresponds to building a vector whose components are the concatenated columns of the density matrix. Under this mapping, the Liouvillian becomes a linear operator acting on the space ${\cal H}\otimes{\cal H}$, defined as
\begin{align}
{\cal L}&=-i\left(I\otimes\hat H-\hat H^T\otimes I\right)\nonumber\\
&-\frac{1}{2}\sum_j\left(I\otimes\hat A_j^\dagger\hat A_j+\hat A_j^T\hat A_j^*\otimes I-2\hat A_j^*\otimes\hat A_j\right)\,,
\label{LindbladVec}
\end{align}
{\color{black}where $I$ denotes the identity matrix.} Here, we will assume that the Hilbert space ${\cal H}$ of the problem has dimension $2^N$. Then the vector $|\rho\rangle$ is defined in a $2^{2N}$-dimensional space and can therefore be encoded onto a $2N$-qubit register. 

The Liouvillian operator for a dissipative system is in general not Hermitian. Its eigenvalues $\lambda_j$ are in general complex, and the nonzero eigenvalues are characterized by $\mathrm{Re}(\lambda_j)<0$. The right and left eigenvectors associated to the zero-eigenvalue (here assumed unique) are respectively the NESS $\hat\rho_{ss}$ and the identity matrix, i.e. ${\cal L}|\rho_{ss}\rangle=0$ and $\langle I|{\cal L}=0$ \cite{Breuer2002,Minganti2018}. The latter relation implies also ${\cal L}^\dagger|I\rangle=0$. We define a Hermitian operator in a $2^{2N+1}$-dimensional space as
\begin{equation}
M = \left(\begin{array}{ll}
0& {\cal L}\\
{\cal L}^\dagger &0
\end{array}\right)\,.
\end{equation}
From the spectral properties of the Liouvillian, we gather that the operator $M$ has two eigenvectors with zero eigenvalue, namely $|\eta_0\rangle=|0\rangle|I\rangle$ and $|\eta_1\rangle=|1\rangle|\rho_{ss}\rangle$. From the assumption that the NESS is unique, we conclude that these will be the only eigenvectors with zero eigenvalue of $M$. More generally, we denote with $|\eta_j\rangle$ and $\varphi_j$ the eigenvectors and the corresponding eigenvalues of $M$. Given the structure of $M$, for each eigenvalue $\varphi_j$ associated to the eigenvector $|\eta_j\rangle=|0\rangle|a_j\rangle+|1\rangle|b_j\rangle$, the vector $|\tilde\eta_j\rangle=|0\rangle|a_j\rangle-|1\rangle|b_j\rangle$ is also eigenvector of $M$ with eigenvalue $-\varphi_j$. The eigenvector condition also implies the relations ${\cal L}^\dagger{\cal L}|b_j\rangle=\varphi_j^2|b_j\rangle$ and ${\cal L}{\cal L}^\dagger|a_j\rangle=\varphi_j^2|a_j\rangle$, thus proving that the eigenvalues of $M$ coincide with those of the operators ${\cal L}^\dagger{\cal L}$ and ${\cal L}{\cal L}^\dagger$. Hence, the values $\varphi_j$ are the singular values of the Liouvillian ${\cal L}$ and, by applying Weyl's inequality for singular values, one can finally show that for the Liouvillian gap $g$ -- i.e. the minimal value of $|\mathrm{Re}(\lambda_j)|$ among all nonzero eigenvalues -- the inequality $g\le\min_{\varphi_j\ne0}|\varphi_j|$ holds. We conclude that the Liouvillian gap $g$ sets a lower bound to the spectral gap of the operator $M$. We will use this property in what follows, when discussing the computational complexity of the quantum algorithm.

To encode the operator $M$, the Liouvillian operator is expressed in terms of a Hermitian and an anti-Hermitian term as ${\cal L}={\cal L}_H-i{\cal L}_A$, with ${\cal L}_H^\dagger={\cal L}_H$ and ${\cal L}_A^\dagger={\cal L}_A$. The operator $M$ is then expressed as
\begin{equation}
M = X\otimes{\cal L}_H+Y\otimes{\cal L}_A\,,
\label{M}
\end{equation}
where $X$ and $Y$ are Pauli operators. If ${\cal L}_H$ and ${\cal L}_A$ can be efficiently encoded in terms of quasi-local operators, then $M$ can also be efficiently encoded. From Eq. (\ref{LindbladVec}),
\begin{align}
{\cal L}_A&=\left(I\otimes\hat H-\hat H^T\otimes I\right)+\frac{i}{2}\sum_j\left(\hat A_j^*\otimes\hat A_j-\hat A_j^T\otimes\hat A_j^\dagger\right)\nonumber\\
{\cal L}_H&=\frac{1}{2}\sum_j\left(\hat A_j^*\otimes\hat A_j+\hat A_j^T\otimes\hat A_j^\dagger-I\otimes\hat A_j^\dagger\hat A_j-\hat A_j^T\hat A_j^*\otimes I\right)\,.
\label{LindbladAH}
\end{align}

\section{Quantum algorithm}

\begin{figure}[ht]
  \centerline{
\Qcircuit @C=1em @R=0.2em {
\lstick{} & \qw \barrier[-7.5mm]{16} & \multigate{5}{H^{\otimes t}} & \qw & \qw & \qw & \qw & \cdots & & \ctrl{11} & \qw & \multigate{5}{FT^\dag} & \qw \barrier[-9mm]{16}  \barrier[7.5mm]{16}     & \multigate{5}{\hspace{-5mm}\metersymb\hspace{-5mm}}     \\
\lstick{} & \qw & \ghost{H^{\otimes t}} & \qw & \qw & \qw & \qw & \cdots & & \qw & \qw &              \ghost{FT^\dag} & \qw &        \ghost{\hspace{-5mm}\metersymb\hspace{-5mm}}              \\
\lstick{} & \qw & \ghost{H^{\otimes t}} & \qw & \qw & \qw & \qw & \cdots & & \qw & \qw &              \ghost{FT^\dag} & \qw &        \ghost{\hspace{-5mm}\metersymb\hspace{-5mm}}              \\
\lstick{|\mathbf{0}\rangle} & \qw & \ghost{H^{\otimes t}} & \qw & \qw & \qw & \qw & \cdots & & \qw & \qw &              \ghost{FT^\dag} & \qw &        \ghost{\hspace{-5mm}\metersymb\hspace{-5mm}}               \\
\lstick{} & \qw & \ghost{H^{\otimes t}} & \qw & \qw & \ctrl{7} & \qw & \cdots & & \qw & \qw &         \ghost{FT^\dag} & \qw &        \ghost{\hspace{-5mm}\metersymb\hspace{-5mm}}              \\
\lstick{} & \qw & \ghost{H^{\otimes t}} & \ctrl{6} & \qw & \qw & \qw & \cdots & & \qw & \qw &         \ghost{FT^\dag} & \qw &        \ghost{\hspace{-5mm}\metersymb\hspace{-5mm}}            \\
\lstick{} & & \\
\lstick{} & & \\
\lstick{} & & \\
\lstick{} & & \\
\lstick{} & & \\
\lstick{} & \multigate{5}{\mathcal{P}} & \qw & \multigate{5}{U^{2^0}} & \qw & \multigate{5}{U^{2^1}} & \qw & \cdots & & \multigate{5}{U^{2^{t-1}}} & \qw & \qw & \qw & \qw & \qw \\
\lstick{} & \ghost{\mathcal{F}} & \qw & \ghost{U^{2^0}} & \qw & \ghost{U^{2^1}} & \qw & \cdots & & \ghost{U^{2^{t-1}}}                             & \qw & \qw & \qw & \qw & \qw \\
\lstick{} & \ghost{\mathcal{F}} & \qw & \ghost{U^{2^0}} & \qw & \ghost{U^{2^1}} & \qw & \cdots & & \ghost{U^{2^{t-1}}}                             & \qw & \qw & \qw & \qw & \qw \\
\lstick{|\mathbf{0}\rangle} & \ghost{\mathcal{F}} & \qw & \ghost{U^{2^0}} & \qw & \ghost{U^{2^1}} & \qw & \cdots & & \ghost{U^{2^{t-1}}}                      & \qw & \qw & \qw & \qw & \qw & \\
\lstick{} & \ghost{\mathcal{F}} & \qw & \ghost{U^{2^0}} & \qw & \ghost{U^{2^1}} & \qw & \cdots & & \ghost{U^{2^{t-1}}}                             & \qw & \qw & \qw & \qw & \qw \\
\lstick{} & \ghost{\mathcal{F}} & \qw & \ghost{U^{2^0}} & \qw & \ghost{U^{2^1}} & \qw & \cdots & & \ghost{U^{2^{t-1}}}                             & \qw & \qw & \qw & \qw & \qw 
}}
  \caption{Quantum circuit of the NESS solver. The three dashed lines indicate the stages where the states $|\psi_1\rangle$, $|\psi_2\rangle$, and $|\psi_3\rangle$, Eqs. (\ref{psi1}-\ref{psi3}), occur. The Quantum circuit of the gate $\cal P$ for the preparation of the initial state $|\xi\rangle$ for the second register is detailed in Fig. \ref{fig:figure2}}
  \label{fig:figure1}
\end{figure}
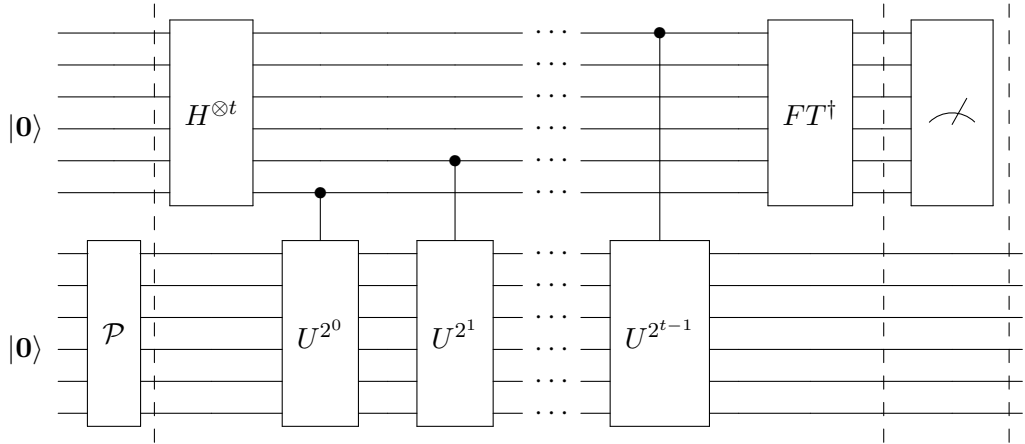

The goal of the algorithm is to prepare an initial state having a large overlap with the degenerate zero eigenspace of $M$. QPE is then applied to estimate the projection of this initial state onto the zero eigenspace. The quantum circuit is illustrated in Fig. \ref{fig:figure1}. It comprises an input-state preparation stage, a QPE stage, and a measurement stage.

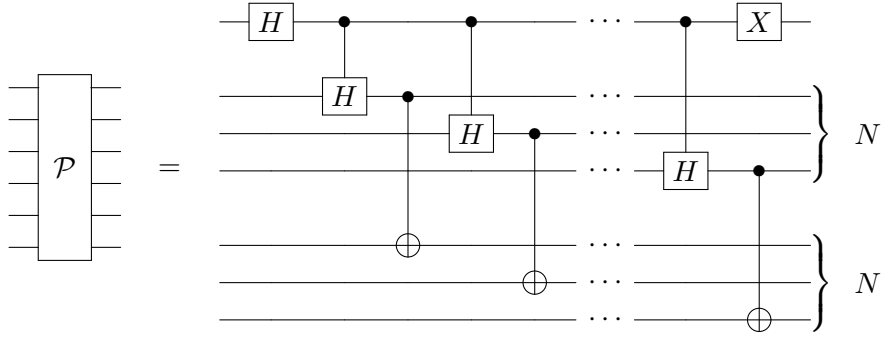
\begin{figure}[ht]
  \centerline{
$\vcenter{
\Qcircuit @C=1em @R=0.2em {
& \multigate{5}{\mathcal{P}} & \qw & \\
& \ghost{\mathcal{F}} & \qw &   \\
& \ghost{\mathcal{F}} & \qw &   \\
& \ghost{\mathcal{F}} & \qw &   \\
& \ghost{\mathcal{F}} & \qw &   \\
& \ghost{\mathcal{F}} & \qw &   
}}=$
$\vcenter{
\Qcircuit @C=1em @R=0.0em @!R {
& & \gate{H} & \ctrl{2} & \qw & \ctrl{3} & \qw & \qw & \cdots && \ctrl{4} & \gate{X} & \qw \\
& & & & & & & & & & & & \\
& & \qw & \gate{H} & \ctrl{4} & \qw & \qw & \qw & \cdots && \qw & \qw & \qw \\
& & \qw & \qw & \qw & \gate{H} & \ctrl{4} & \qw & \cdots && \qw & \qw & \qw &\rstick{N}\\
& & \qw & \qw & \qw & \qw & \qw & \qw & \cdots && \gate{H} & \ctrl{4} & \qw & & \\
& & & & & & & & & & & & \\
& & \qw & \qw & \targ & \qw & \qw & \qw & \cdots && \qw & \qw & \qw \\
& & \qw & \qw & \qw & \qw & \targ & \qw & \cdots && \qw & \qw & \qw &\rstick{N}\\
& & \qw & \qw & \qw & \qw & \qw & \qw & \cdots && \qw & \targ & \qw 
\gategroup{7}{13}{9}{13}{.7em}{\}}
\gategroup{3}{13}{5}{13}{.7em}{\}}
}}$
}
  \caption{Quantum circuit $\cal P$ for the preparation of the initial state $|\xi\rangle$ for the second register. The circuit takes as input the $(2N+1)$-qubit state $|\mathbf{0}\rangle$ and outputs the state $|\xi\rangle$ in Eq. (\ref{xi}). Controlled-Hadamard gates are implemented following the general procedure for arbitrary controlled gates \cite{Nielsen2016,Barenco1995}, and controlled-arbitrary gates are available within current superconducting circuit technology \cite{Lacroix2020}.}
  \label{fig:figure2}
\end{figure}

The QPE algorithm uses two registers. The first one is a $t$-qubit register that will encode a $t$-bit integer estimate of the eigenvalue. The second one encodes the corresponding eigenvector and therefore contains, in our formulation, $2N+1$ qubits. In its standard formulation \cite{Nielsen2016}, QPE takes as an input the state $|\mathbf{0}\rangle|\eta_j\rangle$, where $|\eta_j\rangle$ is an eigenstate of a unitary operator $U$ with eigenvalue $e^{2\pi i\varphi_j}$, assuming $\varphi_j\in[0,1]$. If $2^t\varphi_j$ is a $t$-bit integer, then QPE outputs the state $|\varphi_j\rangle|\eta_j\rangle$. If $2^t\varphi_j$ is not an integer, then QPE will produce with high probability the state $|\tilde\varphi_j\rangle|\eta_j\rangle$, where $2^t\tilde\varphi_j$ is the best $t$-bit integer estimate of $2^t\varphi_j$ from below. 

For the present case, we set $U=e^{2\pi iMt_0}$, where the real parameter $t_0$ is set so that the whole spectrum of $Mt_0$ is included in the $[0,1]$ interval. The algorithm assumes that an oracle circuit is available to efficiently perform controlled-$U^{2^j}$ operations. 
%Given the quasi-local character of $M$ in the computational basis, $U$ can be encoded using algorithms to simulate the time evolution of sparse Hamiltonians \cite{Berry2007,Atia2017,Lloyd1996}, and controlled-$U^{2^j}$ operations are implemented through standard techniques \cite{Nielsen2016}.

Differently from the standard QPE scheme, here the eigenvalue is known and the algorithm is used to compute the corresponding eigenvector. To this purpose, the input state $|\mathbf{0}\rangle|\xi\rangle$ of the QPE stage of the algorithm must be chosen in such a way that $|\xi\rangle$ has a significant overlap with the zero-eigenspace of $M$ spanned by $|\eta_0\rangle$ and $|\eta_1\rangle$. Here we set
\begin{equation}
|\xi\rangle=\frac{|0\rangle|I\rangle+|1\rangle|\mathbf{0}\rangle}{\sqrt{2}}\,.
\label{xi}
\end{equation}
This choice is justified by the procedure to estimate the expectation values of observables that will be introduced below. Insight in the structure of the state $|I\rangle$ shows immediately that it can be expressed as a product of Bell states, where each Bell state involves one qubit from the row index and one from the column index of the density matrix. A circuit on the second register, to prepare the input state (\ref{xi}) when input with the initial state $|\mathbf{0}\rangle$, is shown in Fig. \ref{fig:figure2}. Controlled Hadamard gates can be implemented with a limited number of gates following the prescriptions for arbitrary controlled operations \cite{Nielsen2016,Barenco1995}, and controlled-arbitrary gates are available within current superconducting circuit technology \cite{Lacroix2020}.

Expanding $|\xi\rangle$ on the basis of the eigenvectors $|\eta_j\rangle$ of $M$, the state of the full circuit following the preparation stage is
\begin{align} 
|\psi_1\rangle &= |\mathbf{0}\rangle|\xi\rangle \nonumber\\ 
&=  |\mathbf{0}\rangle\frac{|0\rangle|I\rangle+|1\rangle|\mathbf{0}\rangle}{\sqrt{2}}\nonumber\\
&=  \frac{|\mathbf{0}\rangle|\eta_0\rangle}{\sqrt{2}}+\frac{1}{\sqrt{2}}\sum_{j\ne0}c_{j}|\mathbf{0}\rangle|\eta_j\rangle\nonumber\\
&=  \frac{|\mathbf{0}\rangle|\eta_0\rangle+c_1|\mathbf{0}\rangle|\eta_1\rangle}{\sqrt{2}}+\frac{1}{\sqrt{2}}\sum_{j\ne0,1}c_j|\mathbf{0}\rangle|\eta_j\rangle\,,
\label{psi1}
\end{align}
where we have indicated explicitly the expansion of the state $|1\rangle|\mathbf{0}\rangle$ on the basis of eigenstates $|\eta_j\rangle$ of $M$, and accounted for the fact that states $|1\rangle|\mathbf{0}\rangle$ and $|\eta_0\rangle=|0\rangle|I\rangle$ are orthogonal. The first term of the sum in the last line is the one that will be untouched by the QPE, as both $|\eta_0\rangle$ and $|\eta_1\rangle$ are eigenvectors of $M$ associated to the zero eigenvalue. 

In general, the other eigenvalues of $M$ will be such that $2^t\varphi_j$ is not a $t$-bit integer. Then, the QPE stage of the circuit applied to the state $|\mathbf{0}\rangle|\eta_j\rangle$ will output the state $\sum_{k=0}^{2^t-1}\alpha_k^{(j)}|\mathbf{k}\rangle|\eta_j\rangle$, so that the state of the circuit at the output of the QPE stage is
\begin{equation} 
|\psi_2\rangle = \frac{|\mathbf{0}\rangle|\eta_0\rangle+c_1|\mathbf{0}\rangle|\eta_1\rangle}{\sqrt{2}}+\frac{1}{\sqrt{2}}\sum_{j\ne0,1}\sum_{k=0}^{2^t-1}c_j\alpha_k^{(j)}|\mathbf{k}\rangle|\eta_j\rangle\,,
\label{psi2}
\end{equation}
where
\begin{equation} 
\alpha_k^{(j)}=\frac{1}{2^t}\frac{1-e^{2\pi i(2^t\varphi_j-k)}}{1-e^{2\pi i(\varphi_j-k/2^t)}}\,.
\end{equation}

After the QPE stage, the first register is measured in the computational basis, and the algorithm is successful if the eigenvalue $\mathbf{0}$ is measured. Then, the output state is (up to a normalization constant of order 1)
\begin{equation} 
|\psi_3\rangle = |\mathbf{0}\rangle\left[\frac{|\eta_0\rangle+c_1|\eta_1\rangle}{\sqrt{2}}+\frac{1}{\sqrt{2}}\sum_{j\ne0,1}c_j\alpha_0^{(j)}|\eta_j\rangle\right]\,,
\label{psi3}
\end{equation}
where $c_1\in\mathbb{R}$ because it is the projected amplitude of the density matrix onto its first diagonal matrix element. 
The first term in the sum between brackets in expression (\ref{psi3}) is the result being sought, while the remaining term is an unwanted error proportional to the eigenstates $|\eta_j\rangle$ of $M$ other than those with zero eigenvalues.

\subsection{Success probability and runtime}

From (\ref{psi2}), the probability of measuring zero on the first register is 
\begin{align}
p_0&=\frac{1}{2}\left(1+|c_{0}|^2+\sum_{j\ne0,1}|c_j|^2|\alpha_0^{(j)}|^2\right)\nonumber\\
&=\frac{1+|c_{0}|^2}{2}+p_e\,,
\label{p0}
\end{align}
{\color{black}where $p_e=1/2\sum_{j\ne0,1}|c_j|^2|\alpha_0^{(j)}|^2$ is the probability of an error in the QPE, whereby the zero eigenvalue is measured but the second register is projected onto one of the states $|\eta_j\rangle$ with $j\ne0,1$.} 
As $p_0>1/2$, the number of executions required to successfully measure the zero eigenvalue is $O(1)$ and does not scale with $N$ or $t$.

\begin{figure}[ht]
\centering
\includegraphics[width=\textwidth]{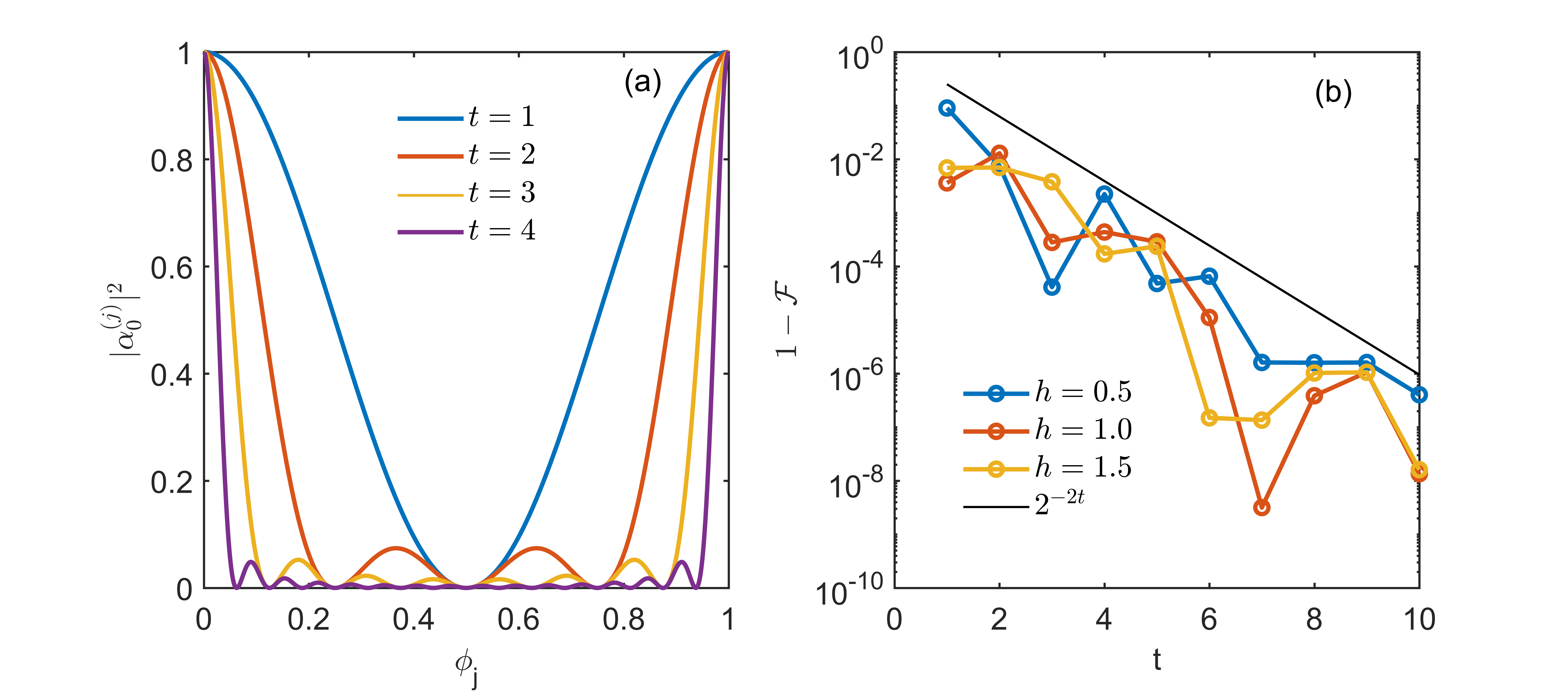}
\caption{(a) The probability $|\alpha_0^{(j)}|^2$ plotted as a function of $\varphi_j$ for different values of $t$. (b) The accuracy $1-{\cal F}$ of the quantum algorithm as a function of $t$, as computed for the simple model of one spin in presence of an external field $h$ along the $x$-direction and relaxation along the $z$-direction.}
\label{fig:figure3}
\end{figure}

In order to estimate the contribution of the spurious term in (\ref{psi3}), it is useful to study the behavior of $|\alpha_0^{(j)}|^2$. Simple algebra leads to
\begin{equation}
|\alpha_0^{(j)}|^2=\frac{1}{2^{2t}}\frac{\sin^2(\pi2^t\varphi_j)}{\sin^2(\pi\varphi_j)}\,.
\label{sinsin}
\end{equation}
The function (\ref{sinsin}) is plotted in Fig. \ref{fig:figure3}(a). {\color{black}As already pointed out, the value of $t_0$ is set in order to have $\varphi_j<1$. The value of $|\alpha_0^{(j)}|^2$ can be large only when $\varphi_j$ approaches an integer value, and the value of $t_0$ can be set such that the only relevant case is when $\varphi_j\sim0$. For $\varphi_j\sim0$ we can expand the denominator and obtain (setting $x=\pi2^t\varphi_j$) $|\alpha_0^{(j)}|^2\simeq\sinc^2(x)$, the squared sine cardinal function. We can use the property $\sinc^2(x)\le x^{-2}$ to set a bound on the error probability
\begin{equation*}
    p_e=\frac{1}{2}\sum_{j\ne0,1}|c_j|^2|\alpha_0^{(j)}|^2\le\frac{1}{2}\sum_{j\ne0,1}\frac{|c_j|^2}{\pi^22^{2t}\varphi_j^2}\le\frac{1}{\pi^2g^22^{2t+1}}\,,
\end{equation*}
where $g$ is the Liouvillian spectral gap and we have used the inequality previously established. Then, in order to have a bound probability of error $p_e<\epsilon_p$, one must set
\begin{equation*}
    t>\left\lceil\log\left(\frac{1}{\sqrt{2}\pi g}\right)+\log\left(\frac{1}{\epsilon_p}\right)\right\rceil\,.
    \label{nqubit}
\end{equation*}
} 

{\color{black}To estimate the gate cost of the algorithm and its scaling with the system size $N$, we first observe that the initial state preparation ${\cal P}$ is made of $2N+2$ gates. The Hadamard stage $H^{\otimes t}$ contributes for $t$ gates, while the QFT stage is implemented with $O(t^2)$ gates if the textbook implementation is used, while approximate versions of the QFT can be implemented with $O(t\log(t))$ gates \cite{Nam2020,Cleve2000}. The oracle part of QPE amounts to the application of a controlled $U^{2^t}=e^{2\pi i 2^tMt_0}$ to the $(2N+1)$-qubit register. The control is implemented at no additional gate cost if a Suzuki-Trotter expansion \cite{Childs2020} is used, as shown explicitly in Appendix A for the case of the dissipative transverse-field Ising model.

A general physical setting is that of $k$-local Hamiltonians $\hat H=\sum_{\gamma}\alpha_{\gamma}\hat H_{\gamma}$ where each $\hat H_{\gamma}$ acts in a non trivial way on $k\sim O(1)$ spins only. We additionally assume that the jump operators $\hat A_j$ in Eq. (\ref{Lindblad}) are also quasi-local operators. Then, the vectorized Liouvillian Eq. (\ref{LindbladVec}) is $k$-local and, from Eq. (\ref{M}), it follows that the effective Hamiltonian $M$ is $(k+1)$-local on an effective system of size $2N+1$. An example is provided by the $3$-local expression of $M$ in the case of the transverse Ising model with local spin relaxation (\ref{Mising}) derived in Appendix A. Optimal strategies for the approximate quantum simulation of Hamiltonian evolution over a time $\tau$ to error $\epsilon_\tau$ have been developed, adopting both Suzuki-Trotter expansion or alternative approaches \cite{Childs2020}. The Suzuki-Trotter approach has the advantage of not requiring ancilla qubits. Assuming a $p_{th}$-order Suzuki-Trotter expansion and the most general $k$-local operator $M$, the best gate cost for the implementation of a $C-e^{i2\pi t_0M\tau}$ gate is estimated \cite{Childs2020} as $O((2N+1)^k\tau^{(1+1/p)}/\epsilon_\tau^{1/p})$ up to the 1-norm of the operator $2\pi t_0||M||$ which in the present case is assumed to be $O(1)$. As noted above, here $\tau=2^t$, which leads us to an estimation of the gate cost $O((2N+1)^k/( g^{1+1/p}\epsilon_\tau^{1/p}\epsilon_p^{1/2})$ for the simulation of the vectorized NESS of a $N$-spin system characterized by a Liouvillian gap $g$, with probability of error $p_e<\epsilon_p$.
The $(2N+1)^k$-dependence is determined by the number of gates required for the most general $k$-local interaction term in the Hamiltonian. However, for models with limited range interactions, the corresponding factor typically depends linearly on $N$ \cite{Tacchino2019}, as for example in the case of the transverse Ising model discussed in Appendix A. The overall gate cost of the present algorithm is therefore
\begin{equation}
N_g=O\left(t+t\log(t)+2N+\frac{(2N+1)^k}{g^{1+1/p}\,\epsilon_\tau^{1/p}\,\epsilon_p^{1/2}}\right)\,, 
\end{equation}
where $\epsilon_\tau$ and $\epsilon_p$ are respectively the Suzuki-Trotter error and the QPE success error. 

Choosing a sufficiently high order $p$, the gate cost approaches $O((2N+1)^k/(g\epsilon_p^{1/2}))$, i.e. linear in the inverse Liouvillian gap and independent of the Suzuki-Trotter error $\epsilon_\tau$. The inverse gap for most model Hamiltonians and couplings to the environment scales as a power law of the system size, except at critical points where it can scale exponentially \cite{Znidaric2015}. The required number of qubits $t$ in the first register, according to Eq. (\ref{nqubit}), is $t\sim O(\log(N))$ in the case of power-law Liouvillian gap, and $t\sim O(N)$ otherwise. Assuming a power-law dependence $1/g=\mathrm{poly}(N)$, the overall gate cost of the algorithm scales with the system size as $O(\mathrm{poly}(N))$, thus achieving a quantum advantage with respect to an exact calculation with a classical algorithm. The scaling with $N$ is instead not favorable whenever the inverse gap scales exponentially with system size. This can happen in some systems when approaching the critical point of a dissipative phase transition, in a way analogous to the simulation of the ground state of a Hamiltonian close to a quantum phase transition using either QPE \cite{Nielsen2016} or a variational quantum eigensolver \cite{McClean2016,Peruzzo2014}.

Previous approaches to the quantum simulation of open quantum systems \cite{Chenu2017,Wang2011,Barreiro2011,Kliesch2011,GarciaPerez2020,Su2020,Hu2020,Sweke2014,Sweke2015,DiCandia2015,Sweke2016,Childs2017,Cleve2017,Liu2020} have focused on algorithms to simulate the time evolution of the system. The scope of the present work is different, in that we propose an algorithm to solve directly the algebraic equation ${\cal L}(\hat\rho_{ss})=0$ that defines the density matrix associated to the NESS. Computing the NESS through time evolution would require a time $T$ long enough to asymptotically approximate the NESS. This time scale is naturally defined as the inverse Liouvillian gap $T\sim1/g$. State-of-the-art approaches \cite{Cleve2017} achieve a gate cost $O(T\mathrm{polylog}(T/\epsilon)$, for a required accuracy $\epsilon$. This is comparable to our approach which scales as $O(T^{1+1/p}/\epsilon^{1/p})$ for a $p_{th}$-order Suzuki-Trotter expansion and approaches a complexity linear in $T$ in the limit of large $p$. The merit of the present approach is that it bypasses the computation of the real time evolution altogether. The transient to the NESS is often characterized by short time features due to the Hamiltonian part of the evolution, that would require a small integration time step, possibly leading to a system dependent computational overhead. Computing directly the density matrix of the NESS avoids this issue and is therefore the method to be preferred if only averaged expectation values of observables on the steady state are needed. A further distinctive feature of the present approach is the fact that its output state encodes directly the elements of the density matrix characterizing the steady state. Most previous approaches instead \cite{Wang2011,Barreiro2011,Kliesch2011,GarciaPerez2020,Su2020,Hu2020,Sweke2014,Sweke2015,Sweke2016,Childs2017,Cleve2017} simulate the quantum channel associated to the open system being studied, typically through appropriate dilation schemes. To estimate the ensemble averaged expectation values of observables, these approaches thus additionally require a possibly large number of runs in order to sample over the statistical ensemble of pure states characterizing the quantum channel. If the goal of the quantum simulation is to estimate ensemble-averaged expectation values, then the present algorithm will likely bring a significant computational advantage. A similar goal is pursued in Ref. \cite{DiCandia2015} where the time evolution of correlators is simulated, although the perturbative scheme used for the dissipative dynamics results in a computational cost scaling exponentially with time, thus inappropriate to extrapolate the NESS in the long-time limit. }

\section{Estimate of expectation values}

Once the first register has been measured in the state $|\mathbf{0}\rangle$, it is possible to estimate quantum mechanical expectation values from the density matrix $\hat\rho_{ss}$. For a given observable $\hat O$, the expectation value is computed as $\langle\hat O\rangle=\mbox{Tr}(\hat O\hat\rho_{ss})/\mbox{Tr}(\hat\rho_{ss})$. As before, we assume that the Hermitian operator $\hat O$ can be efficiently encoded in terms of one- and two-qubit operations, which is generally true for few-point correlations of quasi-local observables. In vectorized form, we have for a general density matrix $|(\hat O\hat\rho)\rangle=I\otimes\hat O|\rho\rangle$, and the super-operator ${\cal O}=I\otimes\hat O$ is also Hermitian. The trace average is then expressed as a matrix element
\begin{equation}
\mbox{Tr}(\hat O\hat\rho_{ss})=\langle I|{\cal O}|\rho_{ss}\rangle=\langle\rho_{ss}|{\cal O}|I\rangle\,.
\label{trace}
\end{equation}

{\color{black}In order to estimate an averaged expectation value (\ref{trace}) from the output state of the algorithm}, we define the Hermitian operator
\begin{equation}
{\cal Q} = X\otimes {\cal O}= \left(\begin{array}{ll}
0& {\cal O}\\
{\cal O} &0
\end{array}\right)\,,
\end{equation}
acting on the $(2N+1)$-qubit second register. Then, 
\begin{equation}
\mathrm{Tr}(\hat O\hat\rho_{ss})=\langle\eta_0|{\cal Q}|\eta_1\rangle=\langle\eta_1|{\cal Q}|\eta_0\rangle\,,
\label{trace2}
\end{equation}
while, due to the structure of the operator $\cal Q$, we have $\langle\eta_0|{\cal Q}|\eta_0\rangle=\langle\eta_1|{\cal Q}|\eta_1\rangle=0$.

The trace average can then be estimated as the expectation value of $I\otimes{\cal Q}$ on the output state $|\psi_3\rangle$
\begin{align}
\langle\psi_3|I\otimes{\cal Q}|\psi_3\rangle&=c_1\mathrm{Tr}(\hat O\hat\rho_{ss})\label{exp}\\
&+\sum_{j\ne0,1}\mathrm{Re}\left[c_j\alpha_0^{(j)}\left(\langle\eta_0|{\cal Q}|\eta_j\rangle+c_1\langle\eta_1|{\cal Q}|\eta_j\rangle\right)\right]+\frac{1}{2}\sum_{j,l\ne0,1}c^*_jc_l\alpha_0^{(j)*}\alpha_0^{(l)}\langle\eta_j|{\cal Q}|\eta_l\rangle\nonumber\,,
\end{align}
where we used the facts that the expectation value of ${\cal O}$ is real-valued, $\langle\eta_0|{\cal Q}|\eta_1\rangle=\langle\eta_1|{\cal Q}|\eta_0\rangle$, and $c_1\in\mathbb{R}$. In particular, by setting $\hat O=I$, the measurement returns the quantity $c_1\mbox{Tr}(\hat\rho_{ss})$, which can be then used to eliminate the factor $c_1$ and obtain the expectation value $\langle\hat O\rangle=\mbox{Tr}(\hat O\hat\rho_{ss})/\mbox{Tr}(\hat\rho_{ss})$. The second and third term on the right hand side of Eq. (\ref{exp}) are the spurious terms resulting from the finite number of qubits $t$ in the QPE. They are respectively of order $O(|\alpha_0^{(j)}|)$ and $O(|\alpha_0^{(j)}|^2)$. By setting $t_0$ as discussed in the previous Section, these terms therefore decrease exponentially in $t$, as also seen in the example presented below.

{\color{black}Eqs. (\ref{trace}) and (\ref{exp}) prove that the problem of computing the trace average of $\hat O$ maps onto that of estimating the expectation value of a Hermitian operator onto the output quantum state of the present algorithm. This estimate is achieved through methods for the quantum measurement of expectation values \cite{Knill2007,Roggero2020}. The accuracy of these methods scales with the number of measurements $N_m$ performed on the output state after, thus requiring $N_m$ successful runs of the algorithm. In order to achieve a given accuracy $\epsilon_m$, a direct operator averaging method requires $N_m=O(1/\epsilon_m^2)$ and no additional gates, while methods based on the QPE can achieve $N_m=O(1/\epsilon_m)$ but at an additional algorithmic cost \cite{Knill2007,Roggero2020}. }

{\color{black}An important question concerns the normalization of the output state encoding the vectorized density matrix, and how it affects the error in estimating expectation values. The appropriately normalized state is written in terms of the elements of the density matrix as $|\rho\rangle=(1/{\cal N})\sum_{jk}\rho_{jk}|j\rangle\otimes|k\rangle$, where ${\cal N}=\sqrt{\sum_{jk}|\rho_{jk}|^2}=\sqrt{P}$ is the Frobenius norm of the density matrix and $P=\mathrm{Tr}(\hat\rho^2)$ the purity of the NESS. If we assume the output state $|\psi_3\rangle$ to be normalized, then the expectation value in Eq. (\ref{exp}) is actually $\langle\psi_3|I\otimes{\cal Q}|\psi_3\rangle=(c_1/P)\mathrm{Tr}(\hat O\hat\rho_{ss})$. In the case of a highly mixed NESS, the large $1/P$-factor amplifies the error in the estimation of $\langle\hat O\rangle$. For a quantum system described in a $2^N$-dimensional Hilbert space, the inequality $1\le1/P\le2^{N}$ holds. The lower bound corresponds to pure states, whereas the upper bound is set by the maximally mixed state corresponding to $\hat\rho=I/2^N$. In practice, physical models of interest usually have poorly entropic steady states, for which $1/P$ is either bound or scales sub-extensively close to a dissipative phase transition \cite{Rota2017,Rota2019}.}

\section{Results}

\begin{figure}[ht]
\centering
\includegraphics[width=\textwidth]{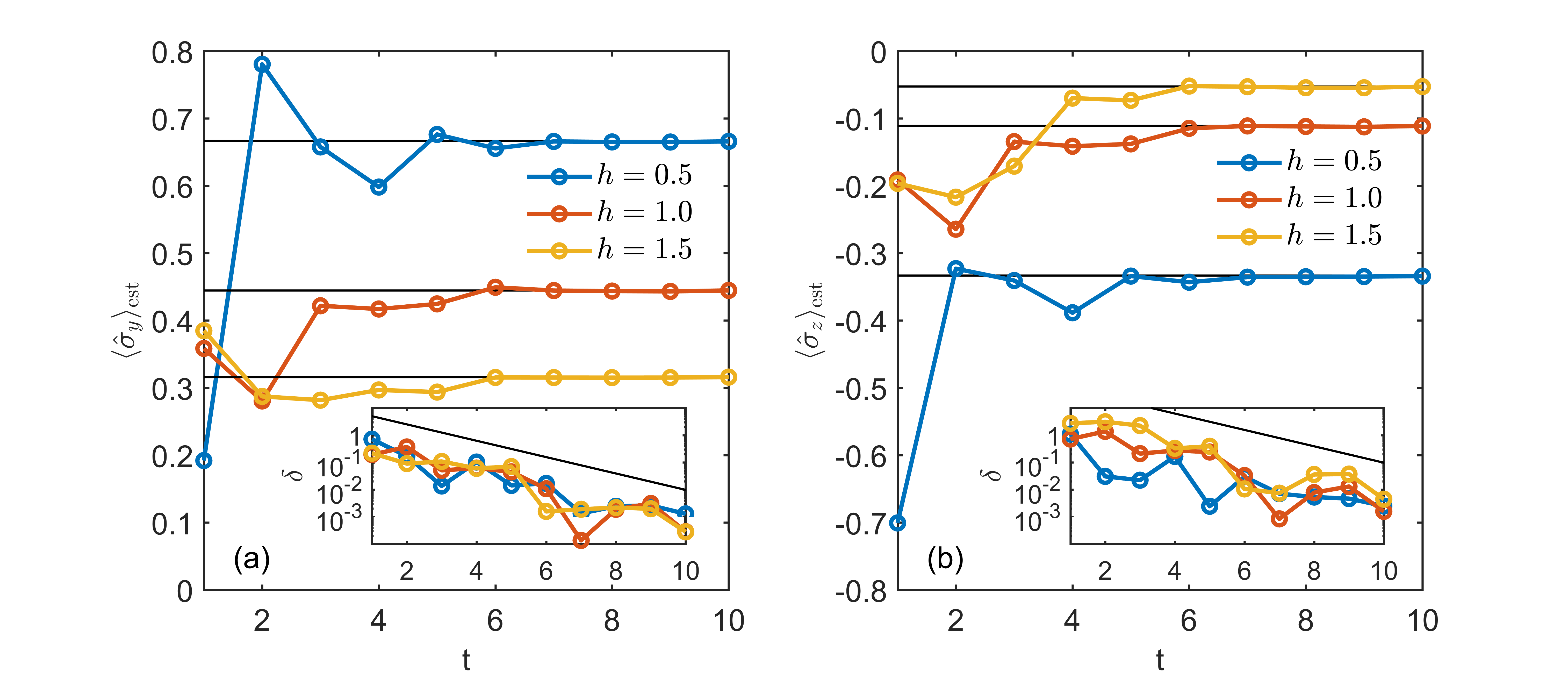}
\caption{Estimated expectation values of (a) $\hat\sigma_y$ and (b) $\hat\sigma_z$, as computed from Eq. (\ref{exp}). The horizontal lines correspond to the values obtained from the exact NESS. The insets show the relative errors $\delta=|\langle\hat\sigma_\alpha\rangle_\mathrm{est}-\langle\hat\sigma_\alpha\rangle|/|\langle\hat\sigma_\alpha\rangle|$ ($\alpha=y,z$) and the black line is a guide to the eye proportional to $2^{-t}$.}
\label{fig:figure4}
\end{figure}

We present here a simple numerical test of the quantum algorithm by simulating the quantum circuit for an elementary model of an open quantum system. We consider a spin 1/2 subject to an external magnetic field $h$ along the $x$ direction and to a decay process along the z-axis. The system Hamiltonian is simply $\hat H=h\hat\sigma_x$ and the density matrix obeys the Lindblad-Von Neumann master equation
\begin{equation}
\dot{\hat\rho}=-i\left[\hat H,\hat\rho\right]-\frac{1}{2}\left(\hat\sigma^{+}\hat\sigma^{-}\hat\rho+\hat\rho\hat\sigma^{+}\hat\sigma^{-}-2\hat\sigma^{-}\hat\rho\hat\sigma^{+}\right)\,.
\end{equation}

The Liouvillian gap for this simple model is independent of $h$ and given by $g=1/2$. We therefore set $t_0=1/5$. The circuit in Figs. \ref{fig:figure1} and  \ref{fig:figure2} was simulated by explicitly computing the corresponding unitary operator, for a number of qubits in the first register up to $t=10$ and three qubits in the second register.

For varying values of $h$ and $t$, we computed the NESS density matrix $\hat\rho_{ss}^{(q)}$ resulting from the quantum algorithm, and compared it to the exact NESS $\hat\rho_{ss}$ by evaluating the fidelity 
\begin{equation}
{\cal F}=\left(\mathrm{tr}\sqrt{\sqrt{\hat\rho_{ss}}~\hat\rho_{ss}^{(q)}\sqrt{\hat\rho_{ss}}}\right)^2\,.
\end{equation}
Results are displayed in Fig. \ref{fig:figure3}(b). The accuracy of the quantum algorithm scales as $2^{-2t }$, as expected from Eqs. (\ref{psi3}) and (\ref{sinsin}). 

In Fig. \ref{fig:figure4} the expectation values of $\hat\sigma_y$ and $\hat\sigma_z$, estimated according to Eq. (\ref{exp}), are plotted as a function of the number of qubits $t$. The corresponding errors with respect to the exact expectation values are plotted in the insets. The leading term linear in $\alpha_0^{(j)}$ in Eq. (\ref{exp}) results in a relative error scaling as $\delta\sim2^{-t}$, again proving the efficiency of the algorithm. 

\section{Conclusions}

We have developed a quantum algorithm for a fault-tolerant quantum computer, that directly estimates the vectorized representation of the NESS density matrix of an open quantum system, without the need to integrate the master equation for the time evolution of the density operator. The algorithm leverages QPE to find the Liouvillian eigenvector associated to the null eigenvalue, thus assuming the existence of a unique NESS. Once this eigenvector is found, we show how to efficiently compute the averaged expectation value of an observable onto the NESS. The scaling of the present quantum algorithm is polynomial in the number of qubits $N$ encoding the quantum degrees of freedom of the system, thus providing quantum advantage when compared to exact diagonalization on classical computer hardware. {\color{black}The algorithm is designed to run on a fault-tolerant quantum computer. As such, we expect it to still execute correctly in a shallow circuit setting where the overall error rate is contained. The question about the scaling of its accuracy, when executed on near-term noisy quantum hardware, remains however open.}

With the availability of scalable fault-tolerant quantum hardware, the present algorithm will enable modeling open quantum systems of unprecedented size, opening a new way in the study of the dissipation and decoherence, including among others, phenomena such as dissipative phase transitions or the influence of the environment on quantum hardware itself.  

\acknowledgments

We acknowledge enlightening discussions with Fabrizio Minganti.

\bibliographystyle{unsrtnat}
\bibliography{qness}

\appendix
{\color{black}
\section{Implementation of the transverse-field Ising model}

We give here a specific circuit implementation of the controlled-$U^{2^j}$ operations in the case of the dissipative transverse-field Ising model, which is currently used as a benchmark for new numerical approaches to open quantum systems \cite{Vicentini2019,DiLuo2020}. We assume that the computational basis encodes spin-1/2 eigenstates of the $Z$ operator. For $N$ spin, the model Hamiltonian is
\begin{equation}
    \hat H = \frac{J}{4}\sum_{\langle j,k\rangle=1}^NZ_jZ_k+\frac{h}{2}\sum_{j=1}^NX_j\,,
    \label{ising}
\end{equation}
where $\langle j,k\rangle$ denotes pairs of indices corresponding to first neighbor sites on the lattice being modeled. Dissipation is described by the jump operator $\hat A_j=\sigma^{(-)}_j=(X_j-iY_j)/2$. Both $\hat H$ and $\hat A_j$ are real-valued in the computational basis, so that $\hat H^T=H$, $\hat A_j^*=\hat A_j$, and $\hat A_j^\dagger=\hat A_j^T=\sigma^{(+)}_j=(X_j+iY_j)/2$.

For clarity, we omit the tensor products in Eq. (\ref{LindbladAH}) from the notation in the formalism that follows. For this, we index the local operators on the left side of the tensor product with indices $j=1,\ldots,N$, and those on the right side with indices $j=N+1,\ldots,2N$. We attribute the index $j=0$ to the additional spin operators appearing on the left side of the tensor product in the definition of $M$, Eq. (\ref{M}). In this way, after simple algebra, the operator $M$ is rewritten as a Hamiltonian for an effective spin model with $2N+1$ spins
\begin{align}
    M &= \frac{J}{4}\sum_{\langle j,k\rangle=1}^NY_0(Z_{j+N}Z_{k+N}-Z_jZ_k)+\frac{h}{2}\sum_{j=1}^NY_0(X_{j+N}-X_j)\nonumber\\
    &+\frac{1}{4}\sum_{j=1}^NY_0(X_jY_{j+N}+Y_jX_{j+N})\nonumber\\
    &+\frac{1}{4}\sum_{j=1}^NX_0(X_jX_{j+N}-Y_jY_{j+N}-Z_{j+N}-Z_j)-\frac{N}{2}X_0\,.
    \label{Mising}
\end{align}
This effective Hamiltonian is $3$-local, i.e. it is written as $M=\sum_{\gamma}\alpha_{\gamma}M_{\gamma}$ where each $M_{\gamma}$ is a product of up to three Pauli operators.

As pointed out above, the controlled unitaries in the QPE part of the algorithm are implemented using the Suzuki-Trotter expansion. As a result, the implementation of the controlled unitary is reduced to the implementation of $C-e^{i\delta M_{\gamma}}$ gates, where $\delta$ is a small real-valued parameter. At the circuit level, this is achieved in terms of CNOTs, $\pm\pi/2$ single-qubit rotations along the $x$ or $y$ axes, and a controlled rotation by an arbitrary angle along the $z$ axis \cite{Tacchino2019}. In Fig. \ref{fig:figure3b} we show the circuits for some selected two- and three-spin controlled operators, while those for the remaining terms occurring in the Suzuki-Trotter expansion can be similarly laid down.

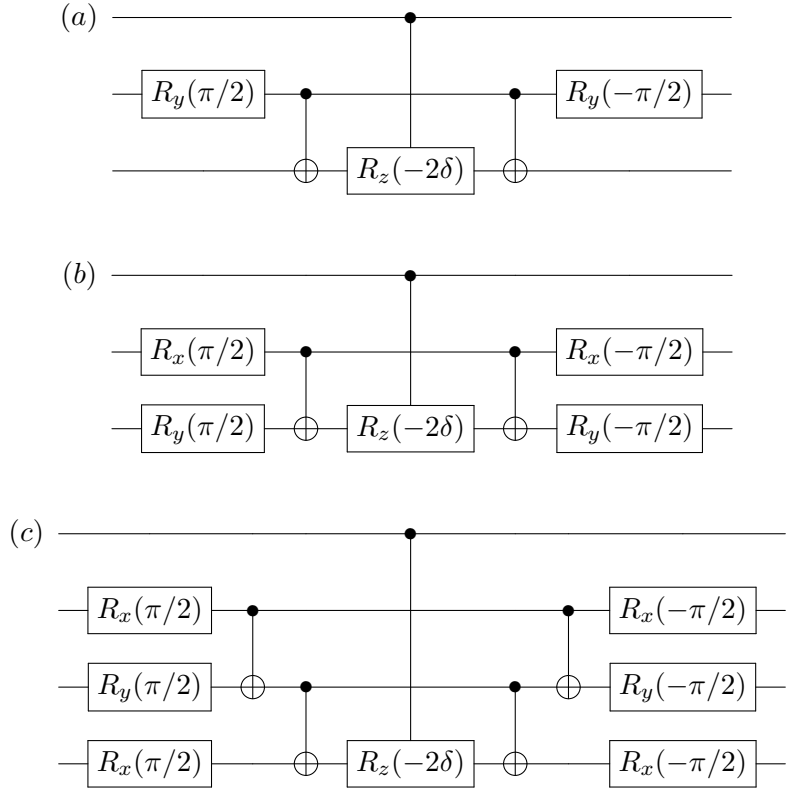
\begin{figure}[ht]
    \centerline{\color{black}
\Qcircuit @C=1em @R=1.0em @!R {
\lstick{(a)}&   \qw    &    \qw   & \ctrl{2} & \qw      & \qw      & \qw  \\
& \gate{R_y(\pi/2)} & \ctrl{1} & \qw      & \ctrl{1} & \gate{R_y(-\pi/2)} & \qw  \\
& \qw      & \targ    & \gate{R_z(-2\delta)} & \targ    & \qw      & \qw  \\
}}
\vspace{1.0cm}
\centerline{\color{black}
\Qcircuit @C=1em @R=1.0em @!R {
\lstick{(b)}&   \qw    &    \qw   & \ctrl{2} & \qw      & \qw      & \qw  \\
& \gate{R_x(\pi/2)} & \ctrl{1} & \qw      & \ctrl{1} & \gate{R_x(-\pi/2)} & \qw  \\
& \gate{R_y(\pi/2)} & \targ    & \gate{R_z(-2\delta)} & \targ    & \gate{R_y(-\pi/2)} & \qw  \\
}}
\vspace{1.0cm}
\centerline{\color{black}
\Qcircuit @C=1em @R=1.0em @!R {
\lstick{(c)}&   \qw             &    \qw   & \qw      & \ctrl{3}             & \qw      & \qw                  & \qw                & \qw  \\
& \gate{R_x(\pi/2)} & \ctrl{1} & \qw      & \qw                  & \qw      & \ctrl{1}             & \gate{R_x(-\pi/2)} & \qw  \\
& \gate{R_y(\pi/2)} & \targ    & \ctrl{1} & \qw                  & \ctrl{1} & \targ                & \gate{R_y(-\pi/2)} & \qw  \\
& \gate{R_x(\pi/2)} & \qw      & \targ    & \gate{R_z(-2\delta)} & \targ    & \qw                  & \gate{R_x(-\pi/2)} & \qw  \\
}}
    \caption{\color{black}Quantum circuits implementing respectively (a) $C-e^{i\delta X\otimes Z}$, (b) $C-e^{i\delta Y\otimes X}$, and (c) $C-e^{i\delta Y\otimes X\otimes Y}$}
    \label{fig:figure3b}
\end{figure}

Assuming to use a first-order Trotter expansion over $r$ Trotter steps, we have
\begin{equation}
    U^j=e^{i2\pi j t_0 M}=\left(e^{i\frac{2\pi j t_0}{r}M}\right)^r\,,
\end{equation}
and
\begin{equation}
e^{i\frac{2\pi j t_0}{r}M}\simeq \prod_\gamma e^{i\frac{2\pi j t_0}{r}\alpha_\gamma M_\gamma}\,.
\label{Trotter1}
\end{equation}
This provides the final prescription to apply the $C-U^j$ as a sequence of gates of the form $C-e^{i\delta M_{\gamma}}$. Based on the circuits shown in Fig. \ref{fig:figure3b}, the gate cost for the implementation of the single Trotter step Eq. (\ref{Trotter1}) is $40N$ single-qubit gates, $42N$ CNOTs and 1 controlled $R_z$ gate. Twice as many gates would be needed if the second order Suzuki-Trotter expansion was employed instead of Eq. (\ref{Trotter1}).

}

\end{document}